\documentclass[runningheads]{llncs}
\usepackage[T1]{fontenc}
\usepackage{fullpage}
\usepackage{graphicx}
\usepackage{booktabs} 
\usepackage{xcolor}
\graphicspath{{./}{./images/}}
\usepackage{amsmath}
\usepackage{algorithm}
\usepackage{algpseudocode}
\usepackage{multirow}
\usepackage{lineno,hyperref}

\modulolinenumbers[1]
\begin{document}
\title{Earthquake Response Analysis with AI}
%
%
\author{Deep Patel\inst{1} \and
Panthadeep Bhattacharjee\inst{2} \and
Amit Reza\inst{3,4} \and
Priodyuti Pradhan\inst{1}}
\authorrunning{Deep et al.}
%
\institute{Department of Computer Science \& Engineering, Indian Institute of Information Technology Raichur, Karnataka - 584135, India. 
\email{\{cs21b1008,prio\}@iiitr.ac.in}
\and
Department of Computer Science \& Engineering, National Institute of Technology, Rourkela, India. 
\email{bhattacharjeep@nitrkl.ac.in}
\and
Space Research Institute, Austrian Academy of Sciences, Schmiedlstrasse 6, 8042 Graz, Austria.
\and 
Nikhef, Science Park 105, 1098 XG Amsterdam, Netherlands. 
\email{amit.reza@oeaw.ac.at}}
\maketitle              
\begin{abstract}
A timely and effective response is crucial to minimize damage and save lives during natural disasters like earthquakes. Microblogging platforms, particularly Twitter, have emerged as valuable real-time information sources for such events. This work explores the potential of leveraging Twitter data for earthquake response analysis. We develop a machine learning (ML) framework by incorporating natural language processing (NLP) techniques to extract and analyze relevant information from tweets posted during earthquake events. The approach primarily focuses on extracting location data from tweets to identify affected areas, generating severity maps, and utilizing WebGIS to display valuable information. The insights gained from this analysis can aid emergency responders, government agencies, humanitarian organizations, and NGOs in enhancing their disaster response strategies and facilitating more efficient resource allocation during earthquake events. 

\keywords{Earthquake \and Disaster Management \and Name Entity Recognition.}
\end{abstract}

\section{Introduction}
A disaster is defined as a significant disruption of a society's ability to function at any scale caused by hazardous events impacting material, economic, environmental, and human losses \cite{1,11}. According to the World Health Organization, earthquakes caused nearly $750,000$ deaths globally between $1998-2017$ \cite{who_earthquake}. Over $125$ million people were affected by earthquakes during this period, meaning they were injured, made homeless, displaced, or evacuated during the emergency phase of the disaster \cite{who_earthquake,9}. Disaster management can be performed in two ways: pre-disaster and post-disaster. Each contains two phases, i.e., four phases in the disaster management life cycle \cite{disaster_management_lifecycle}. The first phase is mitigation, which is responsible for building an early warning system and resilient infrastructure to reduce risk; the second phase is preparedness, which plans for training and emergency plans. The third phase belongs to post-disaster management, i.e., response, responsible for searching, rescuing, and providing essential aid after the disaster. Finally, the fourth phase is the recovery which asses damage and offers financial assistance to restore regular operations. Our work belongs to preparedness, and it will be helpful in the response phases of the disaster management lifecycle.

During natural disasters, millions of people want their voices to be heard and seek updates. Multiple resources are available that provide details of different aspects of the disaster. For effective disaster risk reduction by NGOs and government authorities, it is essential to analyze real-time, ground-level information and use it to inform timely and adaptive responses. Social media data like Twitter serve as valuable sources for several humanitarian objectives, including `situational awareness', especially during disasters \cite{11}. Analyzing this diverse data can offer a comprehensive understanding of the situation \cite{disaster_management_2022,ML_disaster_management_2022}. Despite the increasing reliance on social media for disaster-related insights, existing Twitter analytics methods face limitations \cite{disaster_management_limitations_2022}. 

Early research uses the ML-based model to analyze the tweets for the preparedness phase \cite{NER_earthquake_2021}. A model is trained with some of the collected tweet data sets and tested with the remaining part to find the location information. Finally, create the severity map to provide location information \cite{10,13,20,21,22}. The approach can act as a preparedness phase, but how to use it in the response phase is still being determined. In this article, we proposed a strategy that works as a preparedness phase and can be helpful in the response phase. First, we analyze the last $125$ years of earthquake-prone areas, find all the locations, and use them to pre-train an ML model. After that, during the disaster, the model can take the real-world tweets as test data sets and identify the location to create the severity map. As a case study, we consider the $1st$ January $2024$ Japan earthquake and observe good accuracy from our proposed framework.
\begin{figure}[tbh]
\centering
\includegraphics[height=2.8in, width=3.6in]{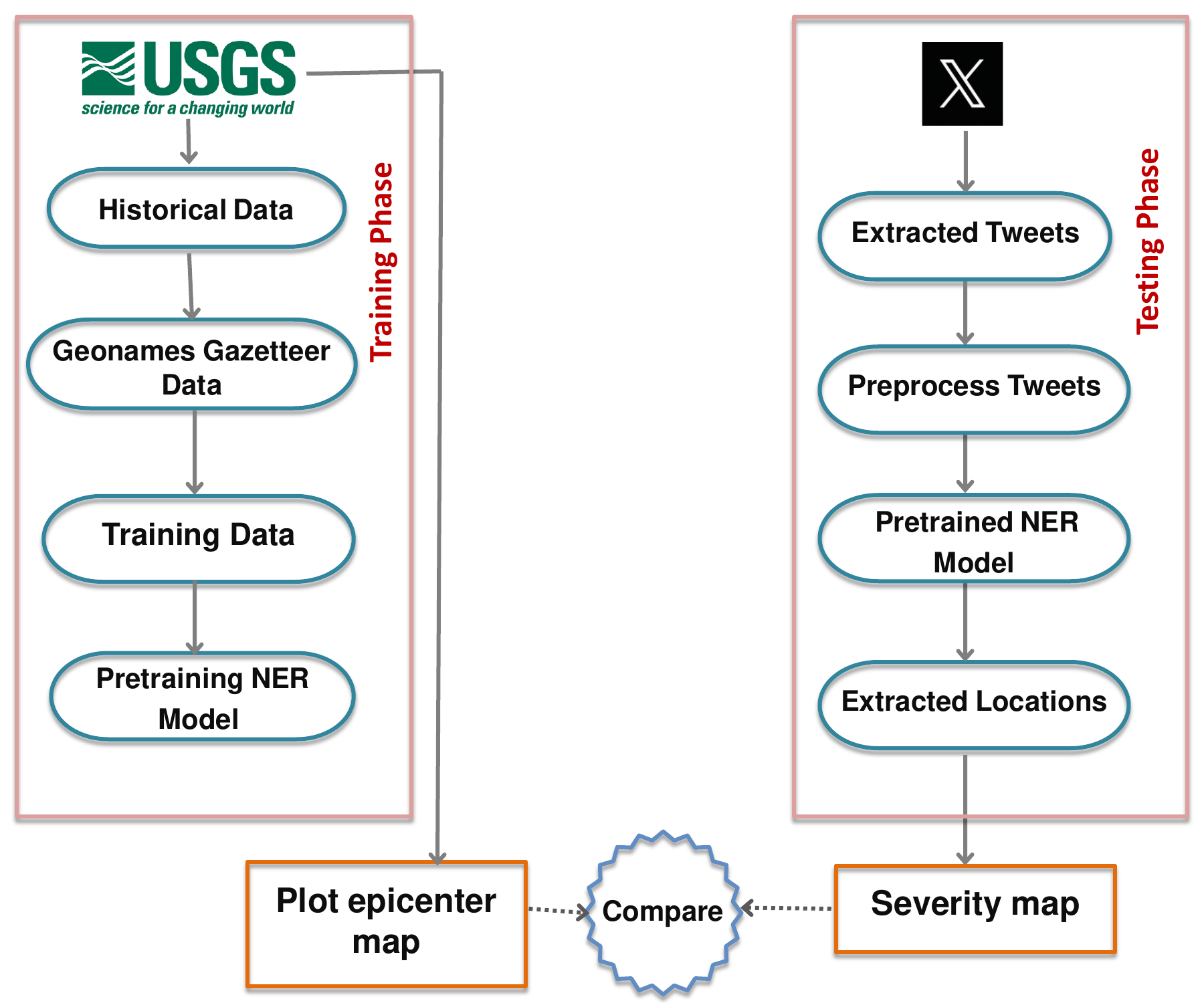}
\caption{Our framework contains the training and testing phase of the AI models. We collect the tweets related to the Kahramanmaraş Earthquake (Turkey-Syria, 2023) for the training data sets from Kaggle. We replaced the locations mentioned in these tweets with all the known locations in Japan. Furthermore, we tag Japan's locations as GPE and disaster-related keywords as DISASTER, leading to the final training data set. Now, we use the pre-trained Name Entity Recognition (NER) model and train it on the training dataset. During the test time, real tweets can be extracted from Twitter, preprocessed, and passed through the pre-trained model for location extraction. Then, they can be mapped to the longitude and latitude for the severity map. Finally, we can compare the severity map with the epicenter map.}
\label{train_test_schematic}
\end{figure}
\begin{figure}[tbh]
\begin{center}
\includegraphics[width = 5.4in, height = 3.5in]{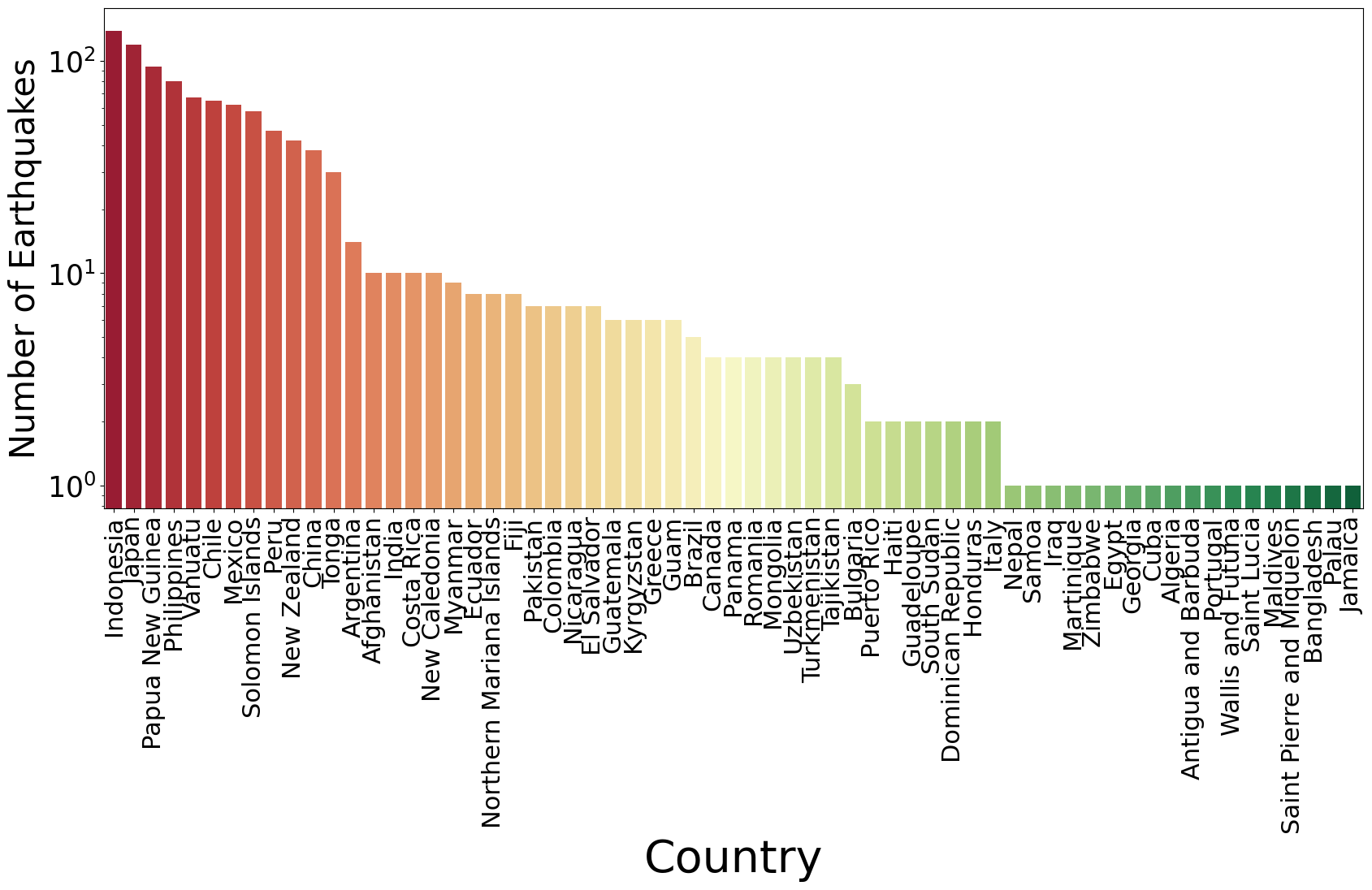}
\caption{It portrays earthquakes worldwide with frequencies over the past $125$ years, with earthquake magnitude on the Richter scale greater than $7.0$. There are a total of $63$ countries.}
\label{earthquake_countrywise}
\end{center}
\end{figure}
\begin{figure}[tbh]\label{tweet_clean}
\centering
\includegraphics[width = 5.5in, height=3.2in]{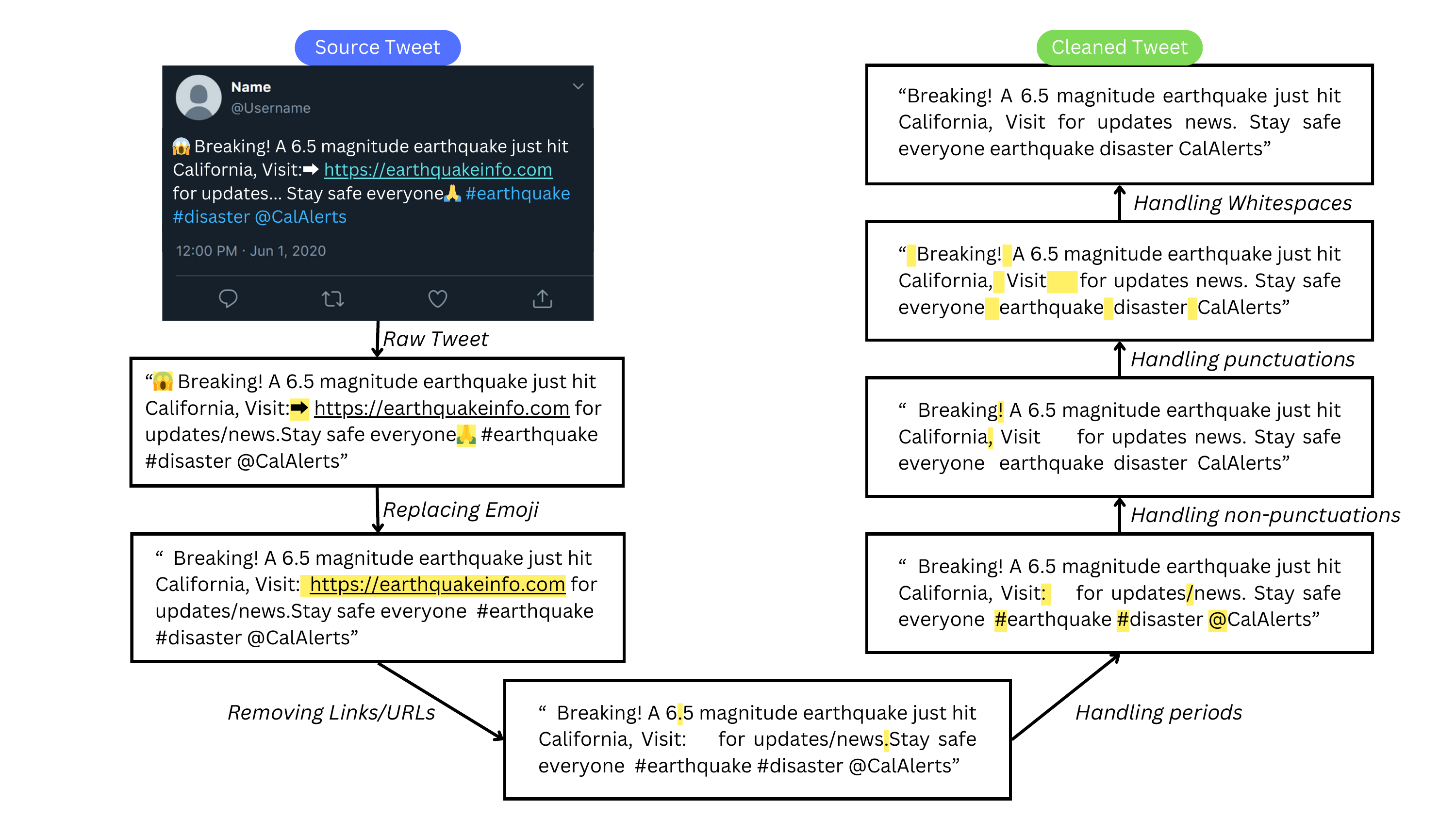}
\caption{Illustrate the Algorithm \ref{algo_preprocessing} on a sample tweet, from collecting raw tweets until we get the cleaned tweet. Each highlighted portion in the text is the target sequence of the respective steps of the process. Non-UTF encoded characters, such as emojis that may not be visible during reading but are still present in the text, are also detected in the initial execution step.}
\label{algo_1_demonstration}
\end{figure}
\begin{figure*}[tbh]
\begin{center}
\includegraphics[width = 6.5in, height = 1.8in]{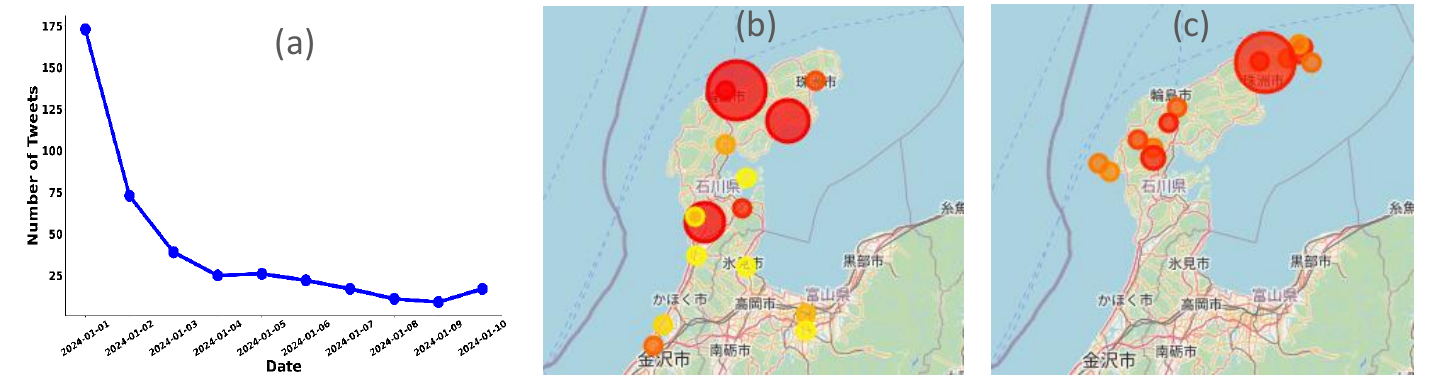}
\caption{(a) Collected tweets during the Japan earthquake on 1st January $2024$. (b) We generate a severity map from locations extracted from disaster tweets from our model during the testing phase. (c) We also plot the epicenter map from seismic data collected from the US Geological Survey (USGS) Earthquake Catalog \cite{USGS}.}
\label{tweets_statistics}
\end{center}
\end{figure*}

\section{Methodology}
The Twitter API allows access to geo-tagged tweets, which offer valuable insights by providing explicit location information. Despite the potential benefits of geo-tagging, various studies highlight that a significant proportion of social media posts lack location data due to user preferences, device limitations, and privacy settings \cite{18}. Further, even after extracting the tweets where most of them are accurately geo-tagged, there are still some vulnerabilities. One significant issue is the geo-tagged location associated with a tweet does not accurately reflect the location referenced in the tweet's content related to disaster areas. For example, a tweet stating, 

\textit{"Help! My nephew is stuck in Tokyo for two days with no food or shelter... \#JapanEarthquake2024"}

\noindent might be geo-tagged as being in Delhi, India. In this scenario, the geo-tag information is misleading because the crisis described in the tweet occurs in Tokyo, Japan, not in Delhi. Therefore, misleading geo-tags can significantly impact the effectiveness of disaster response and resource allocation. Instead of using the geo-tag, we analyze the tweet to extract location information. 

To handle the response phase of the earthquake disaster management, we propose to train an ML model with all the earthquake-prone locations. We collect the locations from the previous history over the last $125$ years from the list of disasters (Fig. \ref{earthquake_countrywise}). We observe that $63$ countries suffered from the earthquakes (Fig. \ref{earthquake_countrywise}). We collect all geolocations for the $63$ countries from gazetteers location databases \texttt{GeoNames}, a comprehensive repository containing structured information about geographical locations worldwide \cite{GeoNames_gazetteer}.
\subsection{Experiment Evaluation}
Our framework consists of five primary components: tweet preprocessing, training data preparation, pretraining an existing model with earthquake-prone areas, location extraction using NER from disaster tweets, and plotting the severity map using extracted geolocations (Fig. \ref{train_test_schematic}). For training purposes, we choose transfer learning and pretrain a name entity recognition (NER) model. The training phase of the custom NER model begins with earthquake location data collection from the \textit{GeoNames} database. 

To create a suitable training dataset for the NER model, we iterate through the entries in the GeoNames data. We extract relevant fields for each entry, such as geopolitical entities (GPE) and their associated geographical identifiers (e.g., longitude and latitude). For particular locations, we create text messages that can mimic a tweet and formatted as a standard (JSON Lines) for the NER task. We make two different data sets for the training.

\noindent {\bf First data set:} Each line of the JSONL file corresponds to a distinct training example, formatted with two primary elements: the text string that includes the geographical name and its corresponding annotations. We have total $80883$ training data sets. For instance, an entry for a place such as "Tokyo" in Japan would be structured in the JSONL file as 
\begin{verbatim}
["Yotsuhama was affected severely by the earthquake.", {"entities": [[0, 9, "GPE"]]}]
\end{verbatim}

\noindent {\bf Second data set:} 
The key idea behind preparing the second dataset is to ensure the model learns the lexical and syntactical patterns associated with named entities in the text and gets realistic tweet data for training. We consider the tweets dataset of the Turkey-Syria 2023 earthquake, which has an initial size of $400k+$ tweets in English, Turkish, and many other languages \cite{16}. We replace the locations mentioned in the tweets with the earthquake-prone locations on which we train our model. To achieve this, we primarily utilize the hashtags in the tweets that contain the locations and other related keywords. Since hashtags are a handy way of grouping and categorizing tweets, anyone searching for a specific topic can find relevant tweets immediately rather than scroll through their X (Twitter) feed \cite{hashtags}. Due to this functionality, we rely on hashtags to gather the locations mentioned in the dataset. Since the dataset contains a specific column for hashtags used in each tweet, we directly use it to gather unique hashtags. If the hashtags are not provided exclusively, the hashtag retrieval from the tweet itself needs to be done before cleaning the tweets. Now that we have gathered the hashtags, We double verify the gathered hashtags using the GeoNames dataset and then using OpenCage geocoding API (since GeoNames data is redundant) and filter the hashtags which are locations \cite{GeoNames_gazetteer,opencage}. Now, we manually identify the locations in tweets and replace them with locations of Japan since we are training only Japan's data (Algorithm \ref{algo_GPE_tagging_training}). For the DISASTER tag, we choose the words related to earthquakes from Table \ref{disaster_keywords} (Algorithm \ref{algo_DISASTER_tagging}). In this way, we prepare the training datasets, which allows the model to learn about all the popular places in Japan (Algorithms \ref{algo_preprocessing}, \ref{algo_GPE_tagging_training}) and disaster-related keywords (Algorithms \ref{algo_DISASTER_tagging}).
\begin{table}[h!]
\centering
\caption{Categorization of Disaster Keywords \cite{Disaster_tagging}.}
\label{disaster_keywords}
\begin{tabular}{@{}lp{10cm}@{}}
\toprule
\textbf{Category} & \textbf{Keywords} \\
\midrule
Core Earthquake & 
earthquake, tremor, aftershock, seismic, fault, epicenter, magnitude, Richter scale, shaking, ground, quake, foreshock, tectonic, plate, shockwave, aftermath, felt, feel \\
\midrule
Descriptive & 
strong, massive, devastating, violent, powerful, intense, mild, deep, surface, shallow \\
\midrule
Damages and Effects & 
damage, damages, damaged, collapse, collapsed, ruins, wreckage, destroyed, cracks, crumbling, aftermath, impact, destruction, disaster, displaced, kill, kills, killed, homeless, injury, injuries, fatalities, debris, rubble, casualties, trapped, death, die, died, wreckage, injured, injury, crashed, crash, blast, blasted \\
\midrule
Responses and Warnings & 
alert, warning, evacuation, rescue, search, emergency, assistance, volunteers, preparedness, shelter, relief efforts, response team \\
\midrule
Measurement and Science & 
Richter, seismograph, seismology, intensity, measurement, scale, USGS, depth, geological, seismometer, seismic \\
\midrule
Natural Disasters and Events & 
tsunami, landslide, fire, eruption, volcano, flood \\
\midrule
Social and Emotional & 
pray, thoughts, fear, panic, trauma, loss, tragedy, devastation, solidarity, support \\
\bottomrule
\end{tabular}
\end{table}
Now, we employ a pre-trained (\texttt{en-core-web-sm}) model from \texttt{spaCy} library and are fine-tuned using location-specific data \cite{en_core_web_sm_models}. During training, the model minimizes prediction errors using labeled datasets, while dropout and regularization techniques prevent overfitting. We choose dropout rates of $0.2$ and batch sizes dynamically increasing from $128$ to $256$ with a $1.3$ growth rate. We use SGD optimizer and learning rate as $1e-5$. The model is configured to undergo training for $40$ epochs, a tiny number. We carefully choose the hyperparameter values used during the training of our NER model to achieve optimal performance. For simplicity, we consider only Japan and denote the populated locations as ${J} = \{j_1, j_2, \dots, j_n \}$ where $n=34k+$ locations. The NER model ($M_{\text{NER}}=\text{en-core-web-sm}$) is trained ($M_{\text{Trained}} \leftarrow \text{Train}(M_{\text{NER}},{J})$) on ${J}$ to identify relevant locations with higher precision. The pretraining enhances the model's ability to identify locations that are not present in general-purpose NER models (Algorithms \ref{algo_preprocessing}, \ref{algo_GPE_tagging_training}, and \ref{algo_DISASTER_tagging}). We can observe the epoch vs. training loss from Fig. \ref{epoch_vs_loss-confusion_matrix}(a).

\begin{table}[ht]
\centering
\begin{minipage}{0.48\textwidth} 
\caption{Test Accuracy on different types of data sets.}
\label{test_results}
\begin{tabular}{|l|l|l|l|}
\hline
{\bf Training} & {\bf Testing} & {\bf Accuracy} & {\bf Entity}\\
\hline
1st dataset & 1st dataset & $\sim 99\%$ & GPE\\
1st dataset & Actual tweets & $\sim 30\%$ & GPE\\
2nd dataset & 2nd dataset & $\sim 99\%$ & GPE\\
2nd dataset & Actual tweet & $\sim 91\%$ & GPE\\
2nd dataset & Actual tweet & $\sim 99\%$ & DISASTER\\
\hline
\end{tabular}
\end{minipage}
\hfill
\begin{minipage}{0.5\textwidth} 
\caption{Classification Report}
\label{classification_report}
\begin{tabular}{lcccccc}
\toprule
\textbf{Label} & \textbf{Precision} & \textbf{Recall} & \textbf{F1-Score} & \textbf{Support} & \textbf{Accuracy} \\
\midrule
DISASTER & 1.00 & 1.00 & 1.00 & 2039 &  1.00 \\
GPE & 1.00 & 0.91 & 0.95 & 1712 & 0.91\\
O & 1.00 & 1.00 & 1.00 & 55331 &1.00 \\
\hline
\end{tabular}
\end{minipage}
\end{table}
\section{Results and Discussion}
This work considers the 1st January $2024$ Japan Earthquake \cite{Japan_earthquake_2024} event as a case study for analyzing the disaster response. We use three keywords to access the earthquake tweets - `japanearthquake', `japanearthquake2024', and `notoearthquake' respectively. The total number of tweets received was $1083$, where $687$ contains some location information, and the rest contain standard situational text.

We analyze the tweets' pattern over time, revealing fluctuations in public awareness and engagement with earthquake events. We observe that the peak is achieved on the day of the disaster itself (Fig. \ref{tweets_statistics}(a)). The number of tweets eventually decreases, indicating a diminishing level of attention and discussion as the situation stabilizes and the immediate crisis subsides. Finally, location detection begins with text preprocessing \cite{NER_earthquake_2021}, ensuring the tweets are cleaned and standardized for the NER model (Figs. \ref{train_test_schematic}, \ref{algo_1_demonstration} and Algorithm \ref{algo_preprocessing}). Then we sequentially tag the GPE (Algorithm \ref{algo_GPE_tagging_testing}) and DISASTER (Algorithm \ref{algo_DISASTER_tagging}) to get the final testing data \( T_{test}^{GD} \). While tagging for GPE in the test data set, we extend the previously defined set of locations \(J\) by adding the popular cities and all the countries in it, thus making the final set \(L\) (Algorithm \ref{algo_GPE_tagging_testing}). This additional data is to test the model's retained knowledge in transfer learning since it was already able to identify the popular cities and countries before fine-tuning.
%
\begin{algorithm}[tbh]
\caption{GPE Tagging of Training Dataset}
\label{algo_GPE_tagging_training}
\begin{algorithmic}
\State \textbf{Input:} Cleaned tweets: \( T_{clean} \) 
\State \textbf{Output:} Training tweets with GPE tags: \( T_{train}^{G} \)
\State \textbf{Steps:}
\State \( J  \gets\) All unique earthquake-prone locations \Comment{Locations of Japan in our case}
\State \( H \gets \{ h \mid h \text{ is a unique hashtag from the } \texttt{hashtag} \text{ column in } T_{clean} \} \) \Comment{Set of distinct keywords extracted from hashtags, potentially indicating locations}
\State \( H_{valid} \gets \emptyset \) \Comment{Initialize a list for verified geographical hashtags}
\For{each hashtag \( h \in H \)}
    \If{\( h \)is in GeoNames data \textbf{and} is validated using OpenCage Geocoding API}
            \State \( H_{valid} \gets H_{valid} \cup \{ h \} \)
        \EndIf
\EndFor
\For{each tweet \( t \in T_{\text{clean}} \)}
    \For{each location \( h \in H_{valid} \)}
        \State Replace \( h \) with \( \#^{|h|} \)
    \EndFor
    \State Replace \( \#^+ \) with \( \# \) \Comment{Simplify consecutive hashes}
    \For{each \( \# \) in \( t \)}
        \State Replace sequentially \( \# \) with \( j \) in \( J \) \Comment{Insert earthquake-prone location names}
    \EndFor
\EndFor
\For{each tweet \( t \in T_{\text{clean}} \)}
    \For{each valid location \( j \in J \) found in \( t \)}
        \State Tag the entity span corresponding to \( j \) with the GPE label
    \EndFor
\EndFor
\State \Return \(T_{\text{train}}^{G}\)
\end{algorithmic}
\end{algorithm}
\begin{algorithm}[tbh]
\caption{GPE Tagging of Test Dataset}
\label{algo_GPE_tagging_testing}
\begin{algorithmic}
\State \textbf{Input:} Cleaned tweets: \( T_{clean} \) 
\State \textbf{Output:} Testing tweets with GPE tags: \( T_{test}^{G} \)
\State \textbf{Steps:}
\State \( L  \gets\) All unique earthquake-prone locations, other popular cities, and countries \Comment{Collected from GeoNames}
\For{each tweet \( t \in T_{\text{clean}} \)}
    \For{each valid location \( l \in L \) found in \( t \)}
        \State Tag the entity span corresponding to \( l \) with GPE label
    \EndFor
\EndFor
\State \Return \(T_{\text{test}}^{G}\)
\end{algorithmic}
\end{algorithm}
\begin{algorithm}[tbh]
\caption{DISASTER Tagging}
\label{algo_DISASTER_tagging}
\begin{algorithmic}
\State \textbf{Input:} GPE-tagged tweet dataset: \( T^{G} \) (either \( T_{train}^{G} \) or \( T_{test}^{G} \))
\State \textbf{Output:} Tweets with GPE and DISASTER tags: \( T^{GD} \) (either \( T_{train}^{GD} \) or \( T_{test}^{GD} \))
\State \( D \gets \) Set of disaster-related keywords
\State \( T^{GD} \gets T^{G} \) \Comment{Initialize output dataset}
\For{each tweet \( t \in T^{GD} \)}
    \For{each keyword \( d \in D \) found in \( t \)}
        \State Tag the entity span of \( d \) with DISASTER label
    \EndFor
\EndFor
\State \Return \( T^{GD} \)
\end{algorithmic}
\end{algorithm}
Finally, we pass the test data sets ($ T_{test}^{GD}$) through the trained model ($M_{\text{Trained}}$) to extract the location information. 
\begin{equation} \nonumber
\text{Entities}(T_i) = M_{\text{Trained}}(T_i)
\end{equation}

These entities include city names, neighborhoods, and geopolitical regions. We can observe that test accuracy varies based on the way we have trained the model on different data sets (Table \ref{test_results}). For the first data sets, accuracy is not good (Table \ref{test_results}). However, accuracy is very high for the second data set (Table \ref{test_results}). Furthermore, we can observe the confusion matrix (Fig. \ref{epoch_vs_loss-confusion_matrix}(b)) and the classification report (Table \ref{classification_report}), which shows our model gives $96\%$ accuracy in predicting the GPE and DISASTER tag in the test data tweets. 

Finally, for each extracted location entity, we can get the coordinates $(lat_{i}, lon_{i})$ from the Geonames database and create the severity map from the extracted locations (Fig. \ref{tweets_statistics}(b)). Finally, we compare the severity map with the extracted location of tweets and the plot of the epicenter and affected areas (Fig. \ref{tweets_statistics}(c)). In the severity map, yellow circles indicate less severity, and red circles indicate higher severity Fig. \ref{tweets_statistics}(b). The severity maps reflect damage distribution and impact across affected regions. In contrast, the epicenter maps pinpoint where the earthquake originated. Despite these differences in focus, the two maps often correlate, with areas closer to the epicenter typically experiencing higher severity levels (Fig. \ref{tweets_statistics}(b,c)). However, discrepancies may arise due to local building resilience, population density, and proximity to fault lines. Despite potential discrepancies, comparing these maps aids in validating response efforts, refining disaster management strategies, and improving preparedness for future seismic events.
\vspace{.1cm}

\noindent {\bf Low accuracy on the first dataset}: As we mentioned previously, the first dataset contains fixed structure sentences with varying location names. The location names used for preparing the dataset contain almost all the locations of Japan. This way, we train our model to make it "aware" of all the locations of Japan that have the potential to be mentioned in disaster-related tweets. Even though we tried to feed all the locations, the model accuracy is 30\%, which is not very significant. After careful observation, we found out that some more frequently occurring locations (around 10-12 unique locations, occurring around 200 times) in testing data were not present in training data, which means they were not initially present in source data, i.e., GeoNames's gazetteer data. After manually adding those locations to the training data, we observed a very minor accuracy boost of 32-33\%, which is not yet efficient. There must be some serious issue in the training dataset that causes this significant accuracy drop. After reviewing the dataset preparation method and studying the NER training pipeline, we noted some potential causes that can hold back the model from achieving higher accuracy. The causes can be summarized as follows: 
\begin{figure}[tbh]
\centering
\includegraphics[width = 6.5in, height = 2.2in]{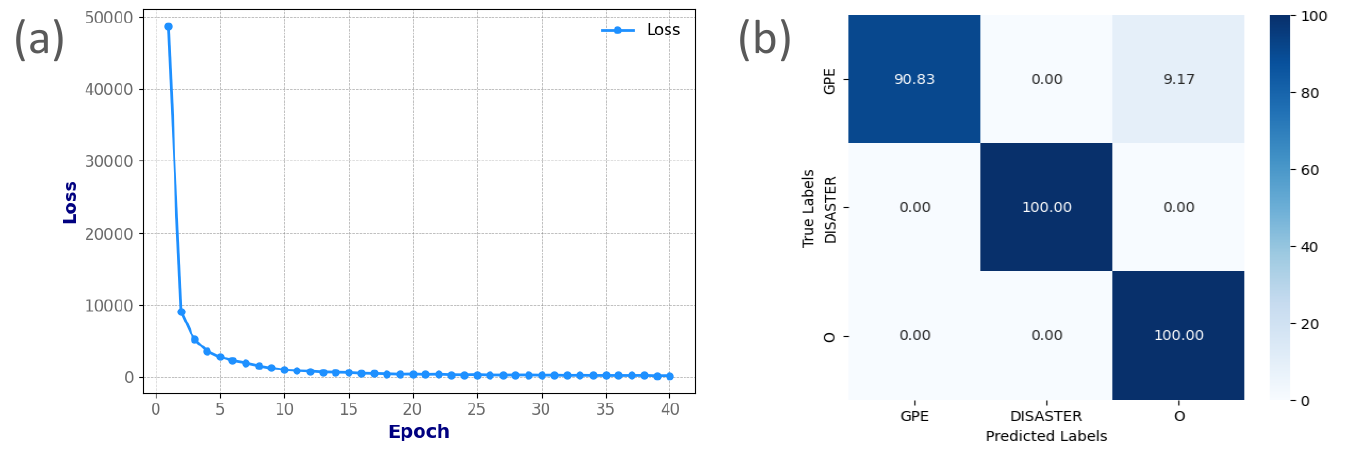}
\caption{NER model on the second dataset. (a) epochs vs. training loss. The curve represents the model's convergence during transfer learning; the loss decreases as the number of epochs increases. (b) The confusion matrix represents the performance of the trained NER model. The model classifies each word of the test dataset into three different entities: GPE (Geopolitical Entity), DISASTER, and O (Other). The `Other' category (O) includes words that do not fall under the GPE or DISASTER entity types. These words represent general terms, non-entity references, or irrelevant information in the context of the classification task.}
\label{epoch_vs_loss-confusion_matrix}
\end{figure}
\begin{enumerate}
\item The main idea behind NER is using the lexical and syntactical properties of the document to detect entities present within the document. The vocabulary and grammatical knowledge both matter while performing NER. For example, \textit{"Apple is planning to launch its new product next month in New York"}. The word "Apple" can be a fruit or company in this sentence. The dictionary meaning will lead to the model not predicting this entity as an ORG. However, knowing the structure and picking up keywords like "planning to launch" and "new product" can lead the model to think that Apple is being referred to as an ORG here. Since we are using the same structure, the model has a high chance of getting overfitted. To see whether our thought process is correct or not. After checking the prediction made by the trained model, we found out that the model marks the first word of each test data tweet as GPE, which is precisely what we were predicting.
\item Another reason for poor performance is that the dataset does not represent real-world tweets, which can potentially cause typing and grammatical errors and other informal text features. Hence, it can fail while dealing with real-world human-generated tweets. The second training dataset helps to resolve the shortcomings of the first data sets and provides high accuracy to the model.
\end{enumerate}

\begin{algorithm}[tbh]
\label{preprocess_algo}
\caption{Preprocessing Dataset}
\label{algo_preprocessing}
\begin{algorithmic}
\State \textbf{Input:} Raw Tweets: \( T \)
\State \textbf{Output:} Cleaned tweets: \(T_{\text{clean}}\)
\State \textbf{Steps:} 
\State \( T_{\text{en}} \gets \emptyset \) \Comment{Initialize an empty set for English tweets}
\For{each tweet \( t \in T \)}
    \If{\( \text{language}(t) == \text{'en'} \)} \Comment{Check if the tweet is in English}
        \State \( T_{\text{en}} \gets T_{\text{en}} \cup \{ t \} \) \Comment{Add the tweet to the English set}
    \EndIf
\EndFor
\State \( T_{\text{unique}} \gets \emptyset \) \Comment{Initialize an empty set for unique tweets}
\For{each tweet \( t \in T_{\text{en}} \)}
    \If{\( t \notin T_{\text{unique}} \)} \Comment{Check whether the tweet is already in the unique set}
        \State \( T_{\text{unique}} \gets T_{\text{unique}} \cup \{ t \} \) \Comment{Add the tweet to the unique set if not already present}
    \EndIf
\EndFor
\State \(S  \gets\) \{',', '?', '!', ';' \}  \Comment{Define a symbol set of comma, question mark, exclamation mark, and semicolon}
\State \( T_{\text{clean}} \gets \emptyset \) \Comment{Initialize an empty set for cleaned tweets}
\For{each tweet \( t \in T_{\text{unique}} \)}
    \State Remove emojis by replacing them with a single whitespace
    \State Remove substrings in \( t \) matching URL or external link patterns
    \For{each character \( c \) in \( t \)}
        \If{\( c \) == '.'}  \Comment{Check whether it is a period}
            \State Add a single whitespace after periods unless the period is part of a number
        \ElsIf{\( c \) is a symbol and \( c \notin S \)} 
            \State Replace \( c \) with a single whitespace  \Comment{Remove unwanted symbols except basic punctuation}
        \ElsIf{\( c \) is a symbol and \( c \in S \)}
            \State Add a single whitespace after \( c \)
        \EndIf
    \EndFor
    \State Normalize whitespace in \( t \) by removing extra spaces, newlines, tabs, etc. 
    \If{all characters in \( t \) are ASCII} \Comment{Double check for non-English tweets}
        \State \( T_{\text{clean}} \gets T_{\text{clean}} \cup \{ t \} \) \Comment{Add cleaned tweet to the set}
    \EndIf
\EndFor
\State \Return \(T_{\text{clean}}\)
\end{algorithmic}
\end{algorithm}


\section{Conclusion}
Understanding the spatial distribution of earthquake impact is crucial for effective disaster response and mitigation. By integrating NER and geolocation techniques, our approach addresses the shortcomings of traditional geo-tagging methods. The proposed framework provides -- (a) a more reliable mapping of earthquake severity, (b) capturing indirect references to affected areas, (c) landmarks, and (d) local events. The fine-tuning of the NER model on disaster-specific locations ensures that the extracted entities are precise and relevant. While our novel framework offers an improved geo-tagging scheme, the effectiveness of the proposed framework depends on the availability and quality of Twitter data. In regions with low Twitter usage or during disasters when user engagement on the platform is minimal, the analysis may yield insufficient information, compromising the framework's utility.

Our current use of a small-size model is just the beginning. As we expand to cover multiple locations, we can scale up to medium or large-size models, enhancing our training capacity and overall effectiveness. This framework can apply to any other disaster by changing the training process of the ML model. As a follow-up to the current work, we aim to apply Large Language Models (LLMs) like BERT and GPT-4 to retrieve real-time information such as the most affected location. The advantage of LLMs is that they can handle large tweet volumes in real time, eventually decreasing the response delay and providing an emergency response quickly. Further, LLMs can handle multilingual tweets and extract more relevant information that could make a broader impact in handling such a situation. 

\vspace{4mm}
\sloppy \noindent \textcolor{blue}{Code and data availability.} The codes and data to reproduce, examine, and improve our proposed model are available in the GitHub repository \url{https://github.com/coderkage/Earthquake-Response-Analysis-with-AI}


\begin{credits}
\subsubsection{\ackname} Author acknowledges the Anusandhan National Research Foundation (ANRF) grant TAR/2022/000657, Govt. of India.

\subsubsection{\discintname}
Authors have no competing interests.

\end{credits}


\begin{thebibliography}{1}
\bibitem{1} Disaster, United Nations Office for Disaster Risk Reduction. \url{https://www.undrr.org/terminology/disaster}

\bibitem{11}Neppalli, Venkata K., Cornelia Caragea, Anna Squicciarini, Andrea Tapia, and Sam Stehle. Sentiment analysis during Hurricane Sandy in emergency response. International journal of disaster risk reduction 21 (2017): 213-222.



\bibitem{who_earthquake} Earthquakes, World Health Organization. \url{https://www.who.int/health-topics/earthquakes\#tab=tab\_1} 

\bibitem{9} Kim, Jooho, and Makarand Hastak. Social network analysis: Characteristics of online social networks after a disaster. International journal of information management 38, no. 1 (2018): 86-96.

\bibitem{disaster_management_lifecycle} Parsons, Sophie, Peter M. Atkinson, Elena Simperl, and Mark Weal. "Thematically analysing social network content during disasters through the lens of the disaster management lifecycle." In Proceedings of the 24th international conference on world wide web, pp. 1221-1226. 2015.

\bibitem{disaster_management_2022}Guha, Sreeparna, Rabin K. Jana, and Manas K. Sanyal. Artificial neural network approaches for disaster management: A literature review. International Journal of Disaster Risk Reduction 81 (2022): 103276.

\bibitem{ML_disaster_management_2022} Linardos, Vasileios, Maria Drakaki, Panagiotis Tzionas, and Yannis L. Karnavas. Machine learning in disaster management: recent developments in methods and applications. Machine Learning and Knowledge Extraction 4, no. 2 (2022).

\bibitem{disaster_management_limitations_2022}Suwaileh, Reem, Tamer Elsayed, Muhammad Imran, and Hassan Sajjad. When a disaster happens, we are ready: Location mention recognition from crisis tweets. International Journal of Disaster Risk Reduction 78 (2022): 103107.

\bibitem{NER_earthquake_2021} Eligüzel, Nazmiye, Cihan Çetinkaya, and Türkay Dereli. Application of named entity recognition on tweets during earthquake disaster: a deep learning-based approach. Soft Computing 26, no. 1 (2022): 395-421.


\bibitem{10} Belcastro, Loris, Fabrizio Marozzo, Domenico Talia, Paolo Trunfio, Francesco Branda, Themis Palpanas, and Muhammad Imran. Using social media for sub-event detection during disasters. Journal of big data 8, no. 1 (2021): 1-22. 

\bibitem{13} Hernandez-Suarez, Aldo, Gabriel Sanchez-Perez, Karina Toscano-Medina, Hector Perez-Meana, Jose Portillo-Portillo, Victor Sanchez, and Luis Javier García Villalba. Using twitter data to monitor natural disaster social dynamics: A recurrent neural network approach with word embeddings and kernel density estimation. Sensors 19, no. 7 (2019): 1746.

\bibitem{20} Ritter, Alan, Sam Clark, and Oren Etzioni. Named entity recognition in tweets: an experimental study. In Proceedings of the 2011 conference on empirical methods in natural language processing, pp. 1524-1534. 2011.

\bibitem{21} Kumar, Abhinav, and Jyoti Prakash Singh. Location reference identification from tweets during emergencies: A deep learning approach. International journal of disaster risk reduction 33 (2019): 365-375.

\bibitem{22} Wintaka, Deni Cahya, Moch Arif Bijaksana, and Ibnu Asror. Named-entity recognition on Indonesian tweets using bidirectional LSTM-CRF. Procedia Computer Science 157 (2019): 221-228.

\bibitem{18} Zook, Matthew, Mark Graham, Taylor Shelton, and Sean Gorman. Volunteered geographic information and crowdsourcing disaster relief: a case study of the Haitian earthquake. World Medical \& Health Policy 2, no. 2 (2010): 7-33.

\bibitem{GeoNames_gazetteer}Ahlers, Dirk. Assessment of the accuracy of GeoNames gazetteer data. In Proceedings of the 7th workshop on geographic information retrieval, pp. 74-81. 2013. \url{https://download.geonames.org/export/dump/}

\bibitem{15} 
 CrisisMMD: Multimodal Crisis Dataset, \url{https://crisisnlp.qcri.org/crisismmd}

\bibitem{16} Turkey and Syria Earthquake Tweets, \url{https://www.kaggle.com/datasets/swaptr/turkey-earthquake-tweets}

\bibitem{opencage} OpenCage Geocoding API \url{https://opencagedata.com/}

\bibitem{Disaster_tagging} Khan, Saad Mazhar, Imran Shafi, Wasi Haider Butt, Isabel de la Torre Diez, Miguel Angel López Flores, Juan Castanedo Galán, and Imran Ashraf. A systematic review of disaster management systems: approaches, challenges, and future directions. Land 12, no. 8 (2023): 1514.

\bibitem{en_core_web_sm_models} Khasanboyevich, Karimov Jasur, and Zokirov Sanjar Ikromjon Ugli. Software Technologies for Research and Development of Linguistic Models. American Journal of Social and Humanitarian Research 3, no. 5 (2022): 314-320.


\bibitem{Japan_earthquake_2024} Suppasri, Anawat, Miwako Kitamura, David Alexander, Shuji Seto, and Fumihiko Imamura. The 2024 Noto Peninsula Earthquake: Preliminary Observations and Lessons to be Learned. International Journal of Disaster Risk Reduction (2024): 104611.

\bibitem{USGS} US Geological Survey (USGS) Earthquake Catalog,
https://www.usgs.gov/programs/earthquake-hazards/earthquakes

\bibitem{hashtags} How to use X (Twitter) Hashtags: The 2025 Guide to Finding and Using the Right Ones \url{https://www.sendible.com/insights/twitter-hashtags}


\end{thebibliography}
\end{document}